\documentclass[
    ,final            
  ]
  {aipproc}

\layoutstyle{8x11single}

\usepackage{amsmath,amssymb,natbib,graphicx}
\usepackage{color}
\usepackage{ulem}

\begin{document}

\title{Engine-driven Relativistic Supernovae as Sources of\\Ultra High Energy Cosmic Rays}


\author{Alak Ray}{
  address={Department of Astronomy and Astrophysics, Tata Institute of Fundamental Research,\\
    1 Homi Bhabha Road, Mumbai 400 005, India}
}

\author{Sayan Chakraborti}{
  address={Department of Astronomy and Astrophysics, Tata Institute of Fundamental Research,\\
    1 Homi Bhabha Road, Mumbai 400 005, India}
}

\begin{abstract}
Understanding the origin of the highest energy cosmic rays
is a crucial step in probing new physics at
energies unattainable by terrestrial accelerators. Their
sources remain an enigma half a century after their discovery.
They must be accelerated in the local universe as otherwise
interaction with cosmic
background radiations would severely deplete the flux of protons
and nuclei at energies above the Greisen-Zatsepin-Kuzmin (GZK) limit.
Hypernovae, nearby GRBs, AGNs and their flares have all been suggested
and debated in the literature as possible sources. Type Ibc supernovae
have a local sub-population 
with mildly relativistic ejecta which are known to be
sub-energetic GRBs or X-Ray Flashes for sometime and more recently as
those with radio afterglows but without detected GRB counterparts, 
such as SN 2009bb.
In this work we measure the size-magnetic field
evolution, baryon loading and energetics of SN 2009bb using its radio
spectra obtained with VLA and GMRT. We show that
the engine-driven SNe lie above the Hillas line and they can explain the characteristics of post-GZK UHECRs.
\end{abstract}

\maketitle


\section{Introduction}
Direct detection of the highest energy cosmic rays (UHECRs) by satellite-borne instruments is infeasible since these particles have such a large energy
and have so a low flux\cite{2007Sci...318..938T}. They are detected\cite{1963PhRvL..10..146L} only by air
showers\cite{1937RSPSA.159..432B} where the Earth's atmosphere acts as the active medium.
UHECRs beyond the GZK limit\cite{1966PhRvL..16..748G,1966JETPL...4...78Z}
have been invoked
to propose tests of known physical laws and symmetries\cite{1999PhRvD..59k6008C}.
Understanding their origin is important for their use as probes of new physics.
But the sources of UHECRs pose a
problem: the magnetic rigidity of these particles are such that
the magnetic fields in our
galaxy are neither strong enough to contain them nor bend them
sufficiently\cite{1963PhRvL..10..146L}. So any galactic origin would reflect the structure of the galaxy; yet among the UHECRs which have
been detected until now no concentration have been found towards the
Milky Way. Hence their sources are anticipated to be extragalactic.
However UHECR protons at energies above 60 EeV can interact with 
Cosmic Microwave Background (CMB) photons via the $\Delta$ resonance.
The cross section of this process is such that only local extragalactic cosmic
ray sources within 200 Mpc of the Earth can contribute significantly to
the flux of UHECRs above the so called GZK
limit\cite{1966PhRvL..16..748G,1966JETPL...4...78Z,2008PhRvL.100j1101A}. Thus,
to explain the 61 detected cosmic rays with energies above the GZK limit, 
many sources are
required of the UHECRs
\cite{2007Sci...318..938T}.
Since particles of such high energy could not have traveled to Earth from
cosmological distances, unless Lorentz invariance breaks down at these
energies\cite{1999PhRvD..59k6008C}, their detection encourages the search for potential
cosmic ray accelerators in the local Universe.
Accordingly nearby GRBs
\citep{1995PhRvL..75..386W,1995ApJ...449L..37M,2006ApJ...651L...5M},
Hypernovae \citep{2007PhRvD..76h3009W,2008ApJ...673..928B},
AGNs \citep{2007Sci...318..938T} and their flares \citep{2009ApJ...693..329F},
have all been suggested and debated in the literature as possible sources.


SNe with relativistic ejecta have been
detected until recently, exclusively through associated Long GRBs like
GRB 980425\cite{1998Natur.395..663K} or its twin GRB 031203.
XRF 060218\cite{2006Natur.442.1014S} associated with SN 2006aj
showed that mildly relativistic SNe are hundred times less energetic but thousand
times more common (in their isotropic equivalent rate, the relevant rate 
for UHECRs
reaching the observer) than classical
GRBs\cite{2006Natur.442.1014S}. Radio follow up of
SNe Ibc have now discovered the presence of an
engine driven outflow from SN 2009bb\cite{2010Natur.463..513S}, without
a detected GRB.
These mildly relativistic SNe, detected either using X-Ray Flashes (XRFs)
or radio afterglows, a subset of SNe Ibc are far more abundant
at low redshifts required for the
UHECR sources, than the classical GRBs. Their mildly  relativistic
nature, makes them have the most favorable combination 
of $\beta/\Gamma\sim1$, unlike
both non-relativistic SNe and ultra-relativistic classical Long GRBs.
In this work we (\citet{2010arXiv1012.0850C} to appear in Nature Comm.),
have measured the size and magnetic field
of the prototypical SN 2009bb at several epochs. 
Such engine-driven supernovae are placed above
the Hillas line and we demonstrate that they may accelerate cosmic rays beyond the GZK threshold. Together with the rates and energetics of such events,
we
establish that they readily explain the post-GZK UHECRs.

\section{Magnetic field evolution with expanding radius}
In order to derive the highest energy upto which these relativistic SNe
can accelerate cosmic rays (see Figure 1), we determine the
evolution of the size and the magnetic field in the blast-wave.
A Synchrotron Self Absorption (SSA) model fits the
initial radio spectrum of SN 2009bb rather well\cite{2010Natur.463..513S}.
The low frequency
turnover defining the spectral peak shifts to lower frequency with
time characteristic of the expansion of the shocked region that
powers the radio emission. 
Radii and magnetic fields can thus be measured 
from VLA and GMRT data at 5 epochs,
for plotting on the Hillas diagram (Figure 1).

\begin{figure}
\includegraphics[width=0.6\textwidth]{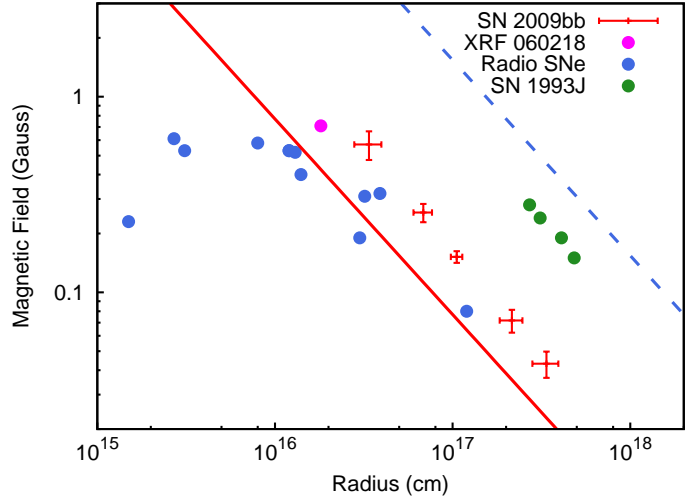}
\caption{\textbf{Hillas Diagram:}
Mildly relativistic sources ($\beta/\Gamma\sim1$) must lie above
the \textit{solid red line}, to accelerate Iron nuclei to 60 EeV
by diffusive shock acceleration\cite{1987PhR...154....1B}, according
to $E_Z\lesssim \beta eZBR / \Gamma$ \cite{2005PhST..121..147W}.
Non-relativistic SNe
($\beta/\Gamma\sim0.05$) must
lie above the \textit{dashed blue line} to reach the same energies.
Radius and magnetic field of SN 2009bb (red crosses, at 5 epochs, determined here
from radio observations with VLA and GMRT assuming equipartition) and
XRF 060218\cite{2006Natur.442.1014S} (magenta ball) lie above the solid red line.
Blue balls denote other\cite{1998ApJ...499..810C} radio SNe with SSA fits.
For SN 1993J only, the magnetic fields are obtained without assuming
equipartition\cite{2004ApJ...612..974C}. All
non-relativistic SNe (blue balls)
including SN 1993J (green balls) lie below the dashed blue line and are unable
to produce UHECRs unlike the mildly relativistic SN 2009bb and XRF 060218 which
lie above the red line. 
}
\end{figure}

With a set of assumptions for the
electron energy distribution and magnetic fields
the radius of the forward shock wave at the time of the synchrotron
self-absorption peak can be written as\cite{1998ApJ...499..810C}
\begin{equation}
R\backsimeq4.0\times 10^{14}\alpha^{-1/19} \left(f\over 0.5\right)^{-1/19}
\left(F_{op}\over {\rm mJy}\right)^{9/19}  \left(D\over {\rm Mpc}\right)^{18/19}
\left(\nu\over 5 {\rm~ GHz}\right)^{-1}  {\rm~cm},
\label{Rp}
\end{equation}
where $\alpha=\epsilon_e/\epsilon_B$ is the ratio of relativistic electron
energy density to magnetic energy density, $f$ is the fraction of the
spherical volume occupied by the radio emitting region, $F_{op}$ is the
observed peak flux, and $D$ is the distance. Using the same variables, the
magnetic field is given by
\begin{equation}
B\backsimeq1.1\alpha^{-4/19} \left(f\over 0.5\right)^{-4/19}
\left(F_{op}\over {\rm mJy}\right)^{-2/19}  \left(D\over {\rm Mpc}\right)^{-4/19}
\left(\nu\over 5 {\rm~ GHz}\right)  {\rm~G}.
\label{Bp}
\end{equation}
The radio spectrum of SN 2009bb at all epochs from discovery
paper (Fig. 2 of ref\cite{2010Natur.463..513S}) and this work,
as obtained from observations using the Very Large Array (VLA) and
the Giant Metrewave Radio Telescope (GMRT), is well fit
by the SSA model. Thus we can 
explicitly
measure the size and magnetic field of a candidate accelerator, instead of
indirect arguments connecting luminosity with the Poynting flux.

SN 2009bb and XRF 060218 can both confine UHECRs and accelerate
them to the highest energies seen experimentally.
At the time of the earliest radio observations\cite{2010Natur.463..513S}
the combination of $\beta/\Gamma\sim1$ for SN 2009bb shows that it
could have accelerated nuclei of atomic number $Z$ to an
energy of $\sim6.5\times Z$ EeV. Thus the source could have
accelerated protons, Neon, and Iron nuclei to 6.4, 64 and 166
EeV respectively. Here the highest energy particles are
likely to be nuclei heavier than protons, consistent with the latest
results indicating an increasing average rest mass of primary UHECRs
with energy\cite{2010PhRvL.104i1101A}. Therefore, our results
support the Auger collaboration's claimed preference of heavier UHECRs 
at the highest energies.

\section{Rates of Engine-driven Supernovae}
We require the rate of relativistic SNe to estimate whether there are enough of them to explain the target objects associated with the $\sim60$ detected UHECRs.
SNe Ibc occur at a rate\cite{1999A&A...351..459C,2004ApJ...613..189D}
of $\sim1.7\times 10^4$ Gpc$^{-3}$ yr$^{-1}$. Their
fraction 
which have relativistic outflows is still somewhat uncertain,
estimated\cite{2010Natur.463..513S} to be around $\sim0.7\%$.
Hence the rate of SN 2009bb-like mildly relativistic
SNe is $\sim1.2\times 10^{-7}$ Mpc$^{-3}$ yr$^{-1}$, 
comparable to the rate of mildly relativistic SNe detected as
sub-energetic GRBs or XRFs of $\sim2.3\times 10^{-7}$ Mpc$^{-3}$ yr$^{-1}$.
This leads to $\sim4$ (or $0.5$) such objects
within a distance of 200 (or 100) Mpc every year.
Since SN 2009bb is still a unique object, only a systematic radio survey
can establish their cosmic rate and statistical properties. 
However, cosmic rays of different
energies have different travel delays due to deflections by magnetic
fields. For a conservative mean delay\cite{2009ApJ...693..329F} of
$\langle \tau_{delay} \rangle \approx 10^5$ yrs
we may receive cosmic rays from any of $4$ (or $0.5$) $\times10^5$ possible
sources at any given time.
Since a direct association between a detected
UHECR and its source is unlikely\cite{2008JCAP...05..006K}, most workers 
have focused on the
constraints\cite{1984ARA&A..22..425H, 2009JCAP...08..026W} placed on
plausible sources. 
Our arguments above show that this new class of objects satisfy all 
such constraints.

Nuclei are subject to photo-disintegration
by interaction with Lorentz boosted CMB photons and can travel upto a distance
of $\sim 100$ Mpc, smaller than but comparable to the GZK
horizon. The local rate of mildly relativistic
SNe is thus high enough to provide enough ($\gg60$) independent sources of
cosmic rays above the GZK limit.
The large $\langle \tau_{delay} \rangle$ implies
that it will not be possible to detect UHECRs from a known relativistic SN,
such as SN 2009bb, within human timescales. However, high
energy neutrinos from photo-hadron interaction at the acceleration site may
be a prime focus of future attempts at detecting these sources with neutrino
observatories like the IceCube.

\section{Energy Injection and Energy Budget}
The required energy injection rate per logarithmic interval in
UHECRs\cite{1995PhRvL..75..386W,2008AdSpR..41.2071B} is
$\Gamma_{inj}=(0.7-20)\times10^{44}$ erg Mpc$^{-3}$ yr$^{-1}$. With the
volumetric rate of mildly relativistic SNe in the local universe,
if all UHECR energy injection is provided by local mildly
relativistic SNe, then each of them has to put in around
$E_{SN}=(0.3-9)\times10^{51}$ ergs. This is comparable to
the kinetic energy in even a normal SN and can easily be supplied
by a collapsar model\cite{1999ApJ...524..262M}.
The mildly relativistic outflow of SN 2009bb 
is in nearly
free expansion for $\sim1$ year. Our measurements of this
expansion 
show \citep{2010arXiv1012.0850C} that this
relativistic outflow, without a detected GRB, is
significantly baryon loaded
and the energy carried by the relativistic baryons is
$E_{Baryons}\gtrsim3.3\times10^{51}$ ergs.

The nearly free expansion of SN 2009bb\citep{2010arXiv1012.0850C} can only
be explained if the mass of the relativistic ejecta is still much larger than
the swept up mass. 
Using the equations for collisional slowdown of the ejecta,
from \citet{1999PhR...314..575P}
and integrating them numerically
(see \citet{2011ApJ...729...57C} for an analytic solution)
from $\gamma_{1}$ to $\gamma_{2}$, the
Lorentz factors at
days 20 and 222 post explosion and substituting for the progenitor mass loss
rate\cite{2010Natur.463..513S}, we solve for the ejecta mass to get
$M_{0}\gtrsim1.4\times10^{-3}M_\odot$. Most of the mass in the relativistic
outflow is due to baryons. The energy associated
with these relativistic protons and nuclei is found to be
$E_{Baryons}\gtrsim3.3\times10^{51}$ ergs.
If $E_{Baryons}$ is distributed equally
over 10 decades in energy, it can account for $\sim0.33\times10^{51}$ ergs
in UHECRs per logarithmic energy interval. With the rate of relativistic 
SNe in the local universe, this is consistent with the
volumetric energy injection rate for UHECRs.

\section{Discussion}
The arrival directions of the Auger events have been claimed to correlate
well with the locations of nearby AGNs\cite{2007Sci...318..938T}; this
suggests that they come from either AGNs or objects with similar
spatial distribution as AGNs. 
The correlation appears weaker in subsequent data\cite{2009Hague}
and the HiRes events\cite{2008APh....30..175T}
do not show such a correlation.
On the other hand, UHECRs correlate well\cite{2008MNRAS.390L..88G}
with the locations of neutral hydrogen (HI) rich galaxies from
the HI Parkes All Sky Survey (HIPASS).
SNe Ibc occur mostly in gas rich star forming spirals. In
particular the 21 cm flux of
NGC3278 (hosting
SN 2009bb) obtained from the HyperLeda database\cite{2003A&A...412...45P}
amounts to
$\sim1.9 \times 10^9 M_\odot$ of HI.
Hence, the observed correlation of UHECR arrival directions with HI selected
galaxies\cite{2008MNRAS.390L..88G} is consistent with our hypothesis. 

We have shown that the newly found subset of nearby
SNe Ibc, with engine-driven mildly relativistic outflows detected as
sub-energetic GRBs, XRFs or solely via their strong radio emission,
can be a source of UHECRs with energies beyond the GZK limit. 
Our study demonstrates for the first time, a new class of objects, which
satisfy the constraints which any proposed accelerator of UHECRs
has to satisfy.


\begin{theacknowledgments}
We thank our collaborators Alicia Soderberg, Abraham Loeb
and Poonam Chandra. We thank
Tsvi Piran and Charles Dermer for discussions at the Annapolis GRB2010 meeting
and the ITC at Harvard Smithsonian Center for Astrophysics for hospitality.
AR also thanks the Physics Dept of West Virginia University for hospitality.
\end{theacknowledgments}



\bibliographystyle{aipproc}   

\bibliography{uhecr}

\IfFileExists{\jobname.bbl}{}
 {\typeout{}
  \typeout{******************************************}
  \typeout{** Please run "bibtex \jobname" to optain}
  \typeout{** the bibliography and then re-run LaTeX}
  \typeout{** twice to fix the references!}
  \typeout{******************************************}
  \typeout{}
 }

\end{document}